# Modification of unconventional Hall effect with doping at the non-magnetic site in a 2D van der Waals ferromagnet

Rajeswari Roy Chowdhury,[1,*] Chandan Patra,[1] Samik DuttaGupta,[2,3,4] Sayooj Satheesh,[1] Shovan Dan,[5] Shunsuke Fukami,[2,3,4,6,7] and Ravi Prakash Singh[1,†]

[1]*Department of Physics, Indian Institute of Science Education and Research Bhopal. Bhopal Bypass Road, Bhauri, Madhya Pradesh 462-066. India*
[2]*Center for Science and Innovation in Spintronics, Tohoku University, 2-1-1 Katahira, Aoba-ku, Sendai 980-8577, Japan*
[3]*Center for Spintronics Research Network, Tohoku University, 2-1-1 Katahira, Aoba-ku, Sendai 980-8577, Japan*
[4]*Laboratory for Nanoelectronics and Spintronics, Research Institute of Electrical Communication, Tohoku University, 2-1-1 Katahira, Aoba-ku, Sendai 980-8577, Japan*
[5]*Department of Physics, The University of Burdwan, Golapbag, Burdwan, West Bengal 713-104. India*
[6]*Center for Innovative Integrated Electronic Systems, Tohoku University, 468-1 Aramaki Aza Aoba, Aoba-ku, Sendai 980-0845, Japan*
[7]*WPI-Advanced Institute for Materials Research, Tohoku University, 2-1-1 Katahira, Aoba-ku, Sendai 980-8577, Japan*

Two-dimensional (2D) van der Waals (vdW) magnetic materials have garnered considerable attention owing to the existence of magnetic order down to atomic dimensions and flexibility towards interface engineering, offering an attractive platform to explore novel spintronic phenomena and functionalities. Understanding of the magnetoresistive properties and their correlation to the underlying magnetic configurations is essential for 2D vdW-based spintronic or quantum information devices. Among the promising candidates, vdW ferromagnet (FM) $Fe_3GeTe_2$ shows an unusual magnetotransport behavior, tunable by doping at the magnetic (Fe) site, and tentatively arising from complicated underlying spin texture configurations. Here, we explore an alternative route towards manipulation of magnetotransport properties of a vdW FM without directly affecting the magnetic site *i.e.*, by doping at the non-magnetic (Ge) site of $Fe_3(Ge,As)Te_2$. Interestingly, doping at the non-magnetic (Ge) site results in an unconventional Hall effect whose strength was considerably modified by increasing As concentration, possibly arising from emergent electromagnetic behavior from underlying complicated spin configurations. The present results provide a possible route to understand the intricate role played by the non-magnetic (Ge) atom towards magnetic properties of vdW FMs, and shows a novel direction towards tailoring of underlying interactions responsible for the stabilization of non trivial spin textures in 2D magnetic vdW materials.

## I. INTRODUCTION

The discovery of two-dimensional (2D) van der Waals (vdW) materials has enunciated a new platform with exciting opportunities, beneficial for the development of new-concept spintronics, magneto-optics, and quantum information processing devices with atomic-scale tunability at reduced dimensions [1-4]. Besides, the stabilization of ferromagnetic and/or antiferromagnetic order down to the monolayer limit in vdW magnets along with its integrability in heterostructures provides an exciting avenue for exploration of fundamental physical concepts and development of novel atomic-scale magnetic devices [5]. Among the different families of 2D vdW ferromagnets (FMs), Cr-based insulating Heisenberg FM $Cr_2Ge_2Te_6$, and metallic $CrI_3$ were initially reported with a bulk transition temperature of ~ 60 K [6,7]. Further investigations resulted in the discovery of itinerant Stoner FM $Fe_3GeTe_2$ (FGT, hereafter), deemed to be prospective owing to its relatively high Curie temperature (~ 220 K) [8-13], large anomalous Hall effect (AHE) [10-14], and significant uniaxial magnetic anisotropy [15]. Structural investigations indicate that bulk FGT can be considered to be built from stacking of $Fe_3Ge$ substructures in a hexagonal crystal structure (space group- $P6_3/mmc$), sandwiched between two Te layers [8,9,16]. For FGT, the existence of spin-orbit coupling effects in the electronic band structure results in strong out-of-plane magnetic anisotropy and topological nodal line-driven large AHE [14], while a frustrated magnetic configuration under applied magnetic fields in bulk and/or few monolayers manifests in the formation of unconventional spin textures and/or skyrmionic lattice structures [17,18]. Besides, these interactions have also been proposed to manifest as an emergent internal magnetic field associated with an unconventional and/or topological Hall effect (THE) from these chiral or topological spin textures [12-19]. Since the possible applications of 2D vdW magnetic materials came into the limelight, investigations on the tuning of properties via electric field [20], strain [21], and mainly chemical doping [22-26] have been a subject of intensive research. In vdW FM $Fe_3GeTe_2$, any variation at the magnetic site (Fe concentration) by doping with a different element leads to transition metal (TM) vacancy-induced profound effects on magnetic properties *viz.* reduction of Curie temperature, magnetic anisotropy strength [24,25]. Magnetotransport studies on Co-doped FGT system have also shown a significant reduction of anomalous Hall resistivity and modification of unconventional and/or topological Hall resistivity associated with changes of the underlying complicated spin configurations [26]. It has been observed that these modifications can be mainly associated with a change in the TM concentration. However, on the lure of understanding the underlying factors determining the exciting properties of FGT and its doped counterparts, manipulation by chemical substitution at the non-magnetic (Ge) site is of utmost interest. In this context, an alternative route towards manipulation of the magnetic properties was shown, where the doping at the non-magnetic (Ge) site of polycrystalline FGT by As results in a significant modulation of the Curie temperature [27]. However, a comprehensive understanding on the effect of doping at the non-magnetic site on the magnetic, magnetotransport and topological properties of a uniaxial vdW FM, either in single-crystalline bulk or in reduced dimensions, have remained elusive. Since single crystals play an important role in presenting a clearer picture of materials' intrinsic properties, growth and exploration of different physical properties of single-crystalline As doped FGT is highly desirable.

[*]rajeswari@iiserb.ac.in
[†]rpsingh@iiserb.ac.in

In this work, we have explored the impact of doping at the non-magnetic (Ge) site of single-crystalline FGT by As and have systematically studied the magnetic, magnetotransport, and topological properties. Magnetic properties with increasing As-doping levels indicate a reduction of $T_C$, possibly attributed to the elongation of the Fe-Fe bonds, consistent with the structural investigations. Magneto-optical investigations reveal a coexisting stripe and bubble-like domain ground state, similar to that observed for Co-doped FGT structures [26]. Magnetotransport measurements indicate the presence of a considerable unconventional contribution to the transverse Hall resistivity, tentatively attributed to the possible existence of underlying complicated emergent spin configurations. Contrary to the previous works which mainly focused on the magnetic (Fe) site, the obtained results demonstrate an unexplored novel pathway towards controlling topological interactions by perturbing the non-magnetic site of a 2D vdW FM, prospective for topological magnetism, and future spintronic devices.

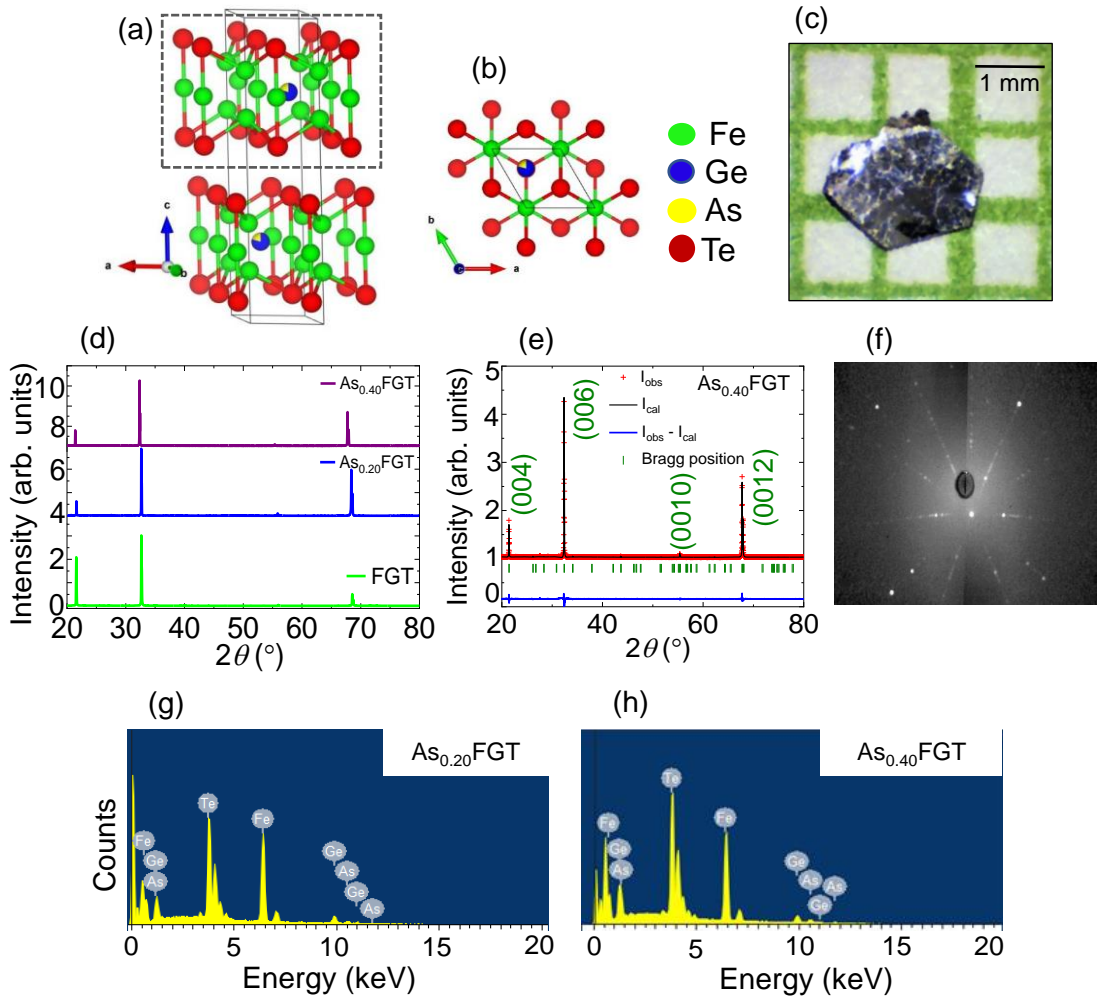

FIG. 1. (a) Crystal structure of $Fe_3(Ge_{1-x}As_x)Te_2$ ($As_x$FGT). The As atoms (yellow circles) are distributed randomly at the Ge site (blue circles). The dashed rectangular area denotes one monolayer of $Fe_3(Ge_{1-x}As_x)Te_2$. (b) Crystal structure as viewed from $ab$-plane with the $c$-axis along out of the plane of the paper. (c) Optical micrograph of $As_{0.20}$FGT single-crystal, utilized in this study. (d) Out-of-plane X-ray diffraction pattern for $As_{0.20}$FGT, $As_{0.40}$FGT, and FGT single-crystalline samples at room temperature. (e) X-ray diffraction of $As_{0.40}$FGT sample with Le Bail fit. (f) Laue diffraction pattern of FGT single crystal. (g), (h) Energy dispersive X-ray analysis (EDX) spectra for $As_{0.20}$FGT and $As_{0.40}$FGT, respectively.

## II. EXPERIMENTAL DETAILS

Single crystalline samples of $Fe_3(Ge_{1-x}As_x)Te_2$ ($As_x$FGT, hereafter) and a reference sample of the parent compound $Fe_3GeTe_2$ (FGT, hereafter) were grown by chemical vapor transport (CVT) method. High purity powder of Fe (5N), Ge (5N), As (5N), and Te (5N) was sealed in a quartz tube along with iodine ($I_2$) as the transport agent and kept at a temperature gradient of 750 °C/700 °C in a two-zone furnace. Shiny silver-colored single crystals were obtained (Fig. 1(c)). Figures 1(a) and (b) show the side and top view of the crystal structure of $As_xFGT$, respectively. Figure 1(d) shows the X-ray diffraction (XRD) pattern of $As_xFGT$ ($x$ = 0.20, 0.40) and FGT, respectively, at room temperature. The x-ray data reveal the samples to be single phase, and the observed Bragg peaks can be indexed with (00*l*) peaks. Besides, it is observed that the diffraction peaks shift towards a lower angle with increasing As concentration ($x$). The Le Bail fit of the XRD data (Fig. 1(e)) shows a reduction of the in-plane lattice parameter $a$ (in $Fe_3Ge_{1-x}As_xTe_2$, $a$ changes from 3.99 Å for $x$ = 0 [8] to 3.92 Å for $x$ = 0.40), while the out-of-plane lattice parameter $c$ increases with increasing As doping ($c$ changes from 16.33 Å for $x$ = 0 to 16.58 Å for $x$ = 0.40), indicating an expansion along the $c$-axis. This change in the lattice parameters is consistent with the experimental results on As doped polycrystalline FGT system [27]. Note that this expansion along the $c$-axis can be possibly attributed to the elongation of the intralayer Fe-Fe bonds. Besides, the $c$-axis expansion by doping at the non-magnetic site is distinctly different from our previous results on magnetic (Fe) site doping, where the effect of doping resulted in a contraction along the $c$-directions [26]. Figure 1(f) shows the Laue diffraction pattern of $As_{0.40}FGT$ confirming the formation of high-quality single crystals, and the inset shows the optical microscope image. The sample compositions and As doping levels ($x$ = 0, 0.20, 0.40) were determined using scanning electron microscope (SEM) equipped with energy dispersive x-ray (EDX) spectrometer (Fig. 1(g), (h)). The magnetic properties were characterized by a vibrating sample magnetometer superconducting quantum interference device (VSM-SQUID) in the temperature range 5-300 K. Magneto-optical investigations were carried out by a polar magneto-optical Kerr effect (MOKE) microscope. Prior to the domain imaging, the samples were freshly cleaved *ex-situ* and transferred into the cryostat. Subsequently, the crystals were cut into a rectangular shape, and a four-probe method was utilized to measure temperature-dependent longitudinal and transverse resistances by physical property measurement system (PPMS). Longitudinal ($\rho_{XX}$) and transverse ($\rho_{XY}$) resistivities were obtained as a function of temperature ($T$) and applied magnetic field ($H$).

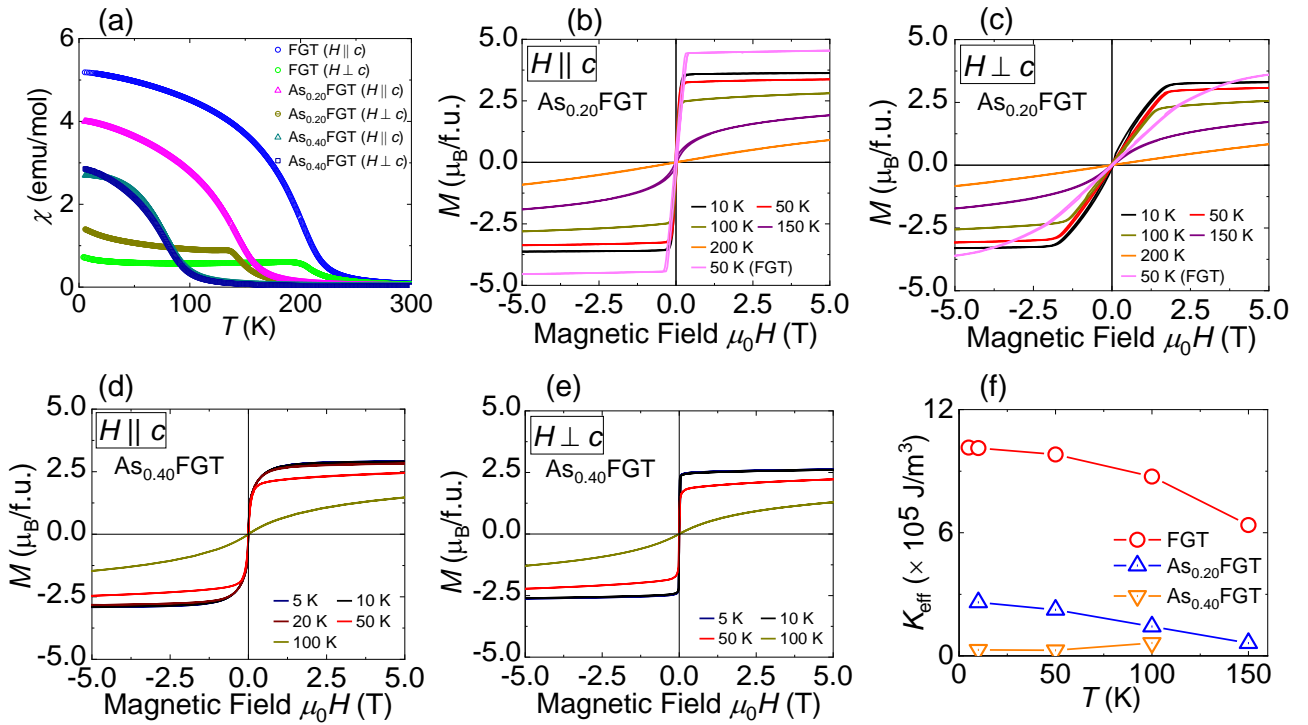

FIG. 2. (a) Magnetic susceptibility ($\chi$) versus temperature ($T$) under magnetic field $\mu_0H$ = 0.5 T, applied parallel to $c$-axis or perpendicular to $c$-axis for $As_xFGT$ and FGT. (b), (c) Field ($H$) dependence of magnetization ($M$) for $As_{0.20}FGT$ and parent FGT single-crystals at various temperature with $H \parallel c$-axis and $H \perp c$-axis, respectively. (d), (e) $M$ versus $H$ for $As_{0.40}FGT$ single-crystal at indicated temperatures for $H \parallel c$-axis and $H \perp c$-axis, respectively. (f) $T$ dependence of effective anisotropy constant ($K_{eff}$) with/without As doping.

## III. RESULTS AND DISCUSSIONS

In this section, we have shown the experimental results and discuss the effect of doping on magnetic and magnetotransport properties of $As_xFGT$ and FGT single crystals. The impact of As-doping on the magnetic properties and the magnetic microstructure have been investigated by magnetization and MOKE techniques. Besides, the effect of As-doping on the anomalous Hall effect, unconventional and/or topological Hall effect, and effective emergent magnetic field were clarified through magnetotransport measurements.

### A. Magnetic properties and microstructure of As-doped FGT (As$_x$FGT)

The effect of doping at the non-magnetic (Ge) site on the magnetic properties of single crystalline FGT were investigated from the temperature ($T$) or magnetic field ($H$) dependence of magnetization ($M$). Figure 2(a) shows $T$ dependence of magnetic susceptibility ($\chi$) for As$_x$FGT and FGT under an applied magnetic field $\mu_0 H$ = 0.5 T ($\mu_0$ = permeability of vacuum), applied parallel ($\parallel$) and perpendicular ($\perp$) to the $c$-axis. From the derivative of $\chi$ along $H \parallel c$ and $\perp$ $c$-axis, we have obtained an averaged Curie temperature ($T_C$) ≈ 200 K (±3 K) for the parent compound FGT, which decreases upon doping to ≈ 145 K (±1 K) and 81 K (±0.5 K), for As$_{0.20}$FGT and As$_{0.40}$FGT, respectively. For FGT, there is a significantly large bifurcation in $M$-$T$ between applied $H \parallel c$ and $\perp$ $c$-axis, consistent with the previous studies and representative of strong magnetic anisotropy of the parent compound [11,12]. However, this bifurcation is considerably weakened with increasing $x$, indicative of a significant reduction of magnetic anisotropy. Figures 2(b)-(e) show the magnetic hysteresis ($M$-$H$) curves for applied $H \parallel c$ and $\perp$ $c$-axes for As$_{0.20}$FGT and As$_{0.40}$FGT. For As$_{0.20}$FGT, the $M$-$H$ curves for $H \parallel c$ attained saturation at a comparatively lower applied field than that for $H \perp c$, confirming the magnetic easy direction along the $c$-axis. On the other hand, for As$_{0.40}$FGT, below 50 K, magnetization curves under applied in-plane magnetic field $H$ ($\perp$ $c$-axis) attain saturated behavior at much lower $H$ compared to that of out-of-plane magnetic fields ($H \parallel c$-axis). However, the coercive fields along $H \parallel c$-axis are smaller than that for $H \perp c$-axis and value of saturation magnetization is also larger along $H \parallel c$-axis, indicating magnetic easy axis along the $c$-direction. The spontaneous magnetization ($M_S$) for the parent compound FGT is the largest at fixed $T$ and progressively decreases with increasing As concentration ($x$). From the isothermal magnetization curves, we have obtained the effective magnetic anisotropy ($K_{eff}$), plotted as a function of $T$ (Fig. 2(f)). A positive sign of $K_{eff}$ within the measured temperature range indicates a magnetic easy-axis along the $c$-axis. Note that the observed suppression of $T_C$ and reduction of $M_S$ and $K_{eff}$ with increasing $x$ show similar trends as our results on doping at the magnetic (Fe) site [26], hinting towards a possible common origin for the observed behavior. Assuming $K_{eff}$ to be equivalent to the uniaxial anisotropy constant ($K_U$), we have calculated the quality factor, $Q = 2K_U/\mu_0 M_S^2$. We have obtained $Q$ (at 10 K) = 4.88, 1.07 for As$_{0.20}$FGT, and As$_{0.40}$FGT, respectively, significantly lower than that obtained for FGT ($Q$ = 11.98 at 10 K) [26].

To further investigate the impact of As-doping on the magnetic microstructure, zero-field cooled (ZFC) MOKE measurements were carried out on As$_{0.20}$FGT, As$_{0.40}$FGT and FGT single crystals. First, a reference image was taken well above $T_C$ of the single crystals. Then, the samples were cooled down to the desired temperatures in the absence of an applied magnetic field. Subsequently, images were acquired at several temperatures in the warming cycle. To enhance the signal-to-noise ratio, the difference between these MOKE images and the reference image was obtained. Figures 3(a)-(c) show zero-field cooled (ZFC) differential MOKE image for As$_{0.20}$FGT (at 120 K), As$_{0.40}$FGT (at 50 K) and parent FGT (at 100 K), respectively. For both As$_{0.20}$FGT and As$_{0.40}$FGT, we have observed a spontaneous generation of stripe-like surface domain structure with alternating contrast corresponding to the magnetization being parallel or antiparallel to the out-of-plane direction (*i.e.*, (00$l$) axis). For As$_{0.20}$FGT, the obtained MOKE images show a stripe-like domain background that embeds rows of circular domains. On the other hand, for As$_{0.40}$FGT, the occurrence of circular domains within this stripe background was not detectable within the limits of resolution (∼ 1 μm) of our MOKE microscope. To obtain the relevant micromagnetic parameters, we utilize a stereological method for the determination of surface domain period ($D$) (see appendix A). From the obtained $D$, we have evaluated domain wall (DW) energy ($\gamma$) = $\frac{DM_S^2}{4\pi\beta}$, where $\beta$ (= 0.31) is a phenomenological parameter dependent on the surface domain arrangement. Subsequently, we have obtained DW width ($\delta = \frac{\pi\gamma}{4K_U}$), and exchange constant ($\varepsilon = \frac{\gamma^2}{16K_U}$). Our experimental results indicate narrow DW widths ($\delta$ ≈ 2-5 nm), comparable to that obtained in ultrathin ferromagnetic heterostructures with perpendicular magnetic anisotropy [28]. The doping at the non-magnetic (Ge) site also resulted in a significant reduction of $\varepsilon$ as compared to that of the parent compound (for instance, $\varepsilon$ (FGT) ≈ 1.08 pJ/m, and that for As$_{0.20}$FGT ≈ 0.11 pJ/m, at 120 K). The present results on the magnetic microstructure and magnetic properties for As$_x$FGT indicate a considerable softening of the ferromagnetic nature, significant than the situation for doping at the magnetic (Fe) site or that of parent FGT [26].

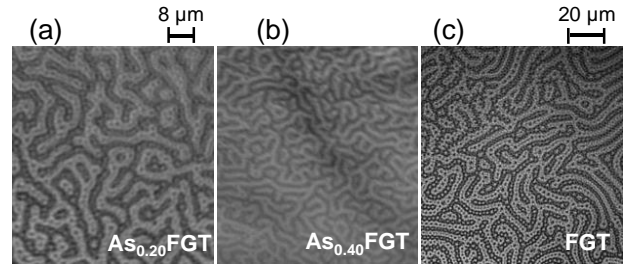

FIG. 3. (a)-(c) Differential p-MOKE image of remnant magnetic state for As$_{0.20}$FGT, As$_{0.40}$FGT, and FGT respectively at 120 K, 50K and 100 K. Black (grey) areas correspond to magnetization pointing in parallel (antiparallel) direction with respect to $c$-axis (out of the plane of paper). The scale bar for the MOKE images of As$_{0.20}$FGT and As$_{0.40}$FGT corresponds to the one above (a), while that for FGT is above (c).

### B. Unconventional Hall effect behavior due to doping at the non-magnetic site

Most of the earlier studies related to the formation of topologically protected non-trivial spin textures and/or emergent electromagnetic behaviors have focused on the presence of Dzyaloshinskii-Moriya interaction (DMI), originating from broken inversion symmetry either in crystallographic lattices lacking inversion sites or in multilayered ferromagnetic heterostructures [29-33]. Recent theoretical results have shown that the presence of magnetic frustration in centrosymmetric FMs (for instance, FGT) can also result in the formation of chiral magnetic structures and/or topological skyrmion lattice, either in bulk or exfoliated heterostructures [34-39]. Subsequent experimental investigations have revealed the existence of square skyrmion lattice in a tetragonal GdRu$_2$Si$_2$ [38], triangular skyrmion-lattice in hexagonal Gd$_2$PdSi$_3$ [37], biskyrmion nano-magnetic domains in hexagonal MnNiGa [39], and unusual Hall effect behavior in FGT, tuneable by doping at the magnetic (Fe) site [26]. In this context, the doping at the non-magnetic (Ge) site of FGT (*i.e.*, without perturbing the TM site) provides an unique opportunity to understand the role of TM towards stabilization of these chiral magnetic structures, and possibly demonstrate an alternative route towards the control of the underlying interactions. Besides, it also provides insights towards the origin of symmetry-breaking chiral interactions in a centrosymmetric structure, providing an unique playground for the investigation of novel interactions in reduced dimensions. To get a deeper understanding concerning the underlying physics, we investigated the magnetotransport properties of parent centrosymmetric FGT and As$_x$FGT single crystals. We expect that any change in the scalar spin chirality associated with the modification of unconventional spin textures as a result of As doping manifests in emergent magnetic field ($B_Z^{\text{eff}}$), resulting in topological Hall and/or unconventional Hall effect behaviors. In a practical scenario, the evaluation of the strength of this unconventional Hall resistivity and $B_Z^{\text{eff}}$ are also essential as it is linked to the density of the underlying skyrmionic and/or other topological spin textures. Figure 4(a) shows the schematic diagram of the measurement configuration utilized in this work. An applied dc $I$ ($\perp$ $c$-axis) of magnitude 10 mA was passed through the single-crystals along the $x$-direction (*i.e.*, along crystal $ab$ plane). The resulting change in $\rho_{XY}$ under simultaneous application of external $H$ (along $z$ or $x$-directions) and $I$ were obtained by measuring voltage drop along the $y$-direction. For applied $H_Z$ ($\parallel$ $c$-axis), we have observed sizable Hall resistance, which can be attributed to AHE in both the crystals [12,13], possibly originating from topological nodal lines in the band structure with the magnetization pointing along the easy axis [14] (Fig. 4(b)). Two characteristic parameters, *viz.* anomalous Hall conductivity (AHC) and the anomalous Hall angle (AHA) are evaluated as $\sigma_{XY}^A = \rho_{XY}/(\rho_{XX}^2 + \rho_{XY}^2)$, and $\theta_{\text{AHA}} = \sigma_{XY}^A/\sigma_{XX}$, respectively, where $\sigma_{XX}$ denotes longitudinal conductivity. $\theta_{\text{AHA}}$ measures the relative contribution of the anomalous Hall current with respect to the charge current amounting to $\theta_{\text{AHA}} \approx 0.05$ (at 10 K) for As$_{0.20}$FGT, which reduces to $\approx 0.01$ for As$_{0.40}$FGT. For FGT, the estimated values of $\theta_{\text{AHA}} \approx 0.06$ (at 100 K) are consistent with the earlier reports, as well [14,40]. The large $\theta_{\text{AHA}}$ for As$_{0.20}$FGT and FGT compared to common ferromagnets ($\theta_{\text{AHA}} \leq 0.02$) indicates the possibility of significant topological contributions in the band structure near Fermi level. Furthermore, $\sigma_{XY}^A$ for FGT and its As-doped counterparts are virtually independent of $\sigma_{XX}$, indicating an intrinsic Berry curvature origin of AHE (see appendix A). Note that this situation is different from that of the doping (or Fe-deficiency) at the magnetic site, where the experimental results were mainly interpreted to arise from a shifting of the spin-orbit coupled electronic bands away from the Fermi level [41]. For our case, the magnetic (Fe) site concentration is relatively constant for different As-doping levels ($x$), pointing to the possibility of an important role played by the non-magnetic atom towards the observed behavior. Besides, $\rho_{XY}$ measurements also manifest in low magnitudes of coercive field and remanence to saturation ratios, confirming the soft ferromagnetic nature, observed from magnetization measurements. On the other hand, for the applied field along $x$-direction ($H_X \parallel I \perp c$-axis), we have observed a reduction of $\rho_{XY}$ magnitude along with the emergence of a prominent cusp-like feature (Fig. 4(d),(e)), whose position shifts to lower field values with increasing $x$, compared to the parent compound FGT. For FGT, the cusp-like peak appears at $\mu_0 H \approx 3$ T [26], which changes to $\approx 1.2$, and 0.6 T, for As$_{0.20}$FGT and As$_{0.40}$FGT, respectively (at 100 K). To examine the strength of the underlying emergent magnetic field ($B_Z^{\text{eff}}$), and to establish a clear picture of the unusual magnetotransport behaviors as a function of As-doping, we have examined the $T$ dependence of $\rho_{XY}$ versus $H_X$ (Fig. 4(d),(e)). The cusp-like behavior in $\rho_{XY}$ persists over the entire temperature range, and an increase of $T$ results in a decrease of $\rho_{XY}$. To extract the unconventional contribution ($\rho_{XY}^U$) from total Hall resistivity, we consider [12,19,26]

$$\rho_{XY} = \mu_0 R_0 H + S_A \rho_{XX}^2 M + \rho_{XY}^U, \qquad (1)$$

where, $\rho_{XX}$ is the longitudinal resistivity of the single crystal (see appendix B), $S_A$ is the field independent coefficient to the anomalous Hall resistivity determined from fitting, and $R_0$ is the ordinary Hall coefficient. Solid lines in Fig. 4 (d), (e) denotes the contributions from ordinary and anomalous Hall contribution which corresponds to the first two terms of Eq.

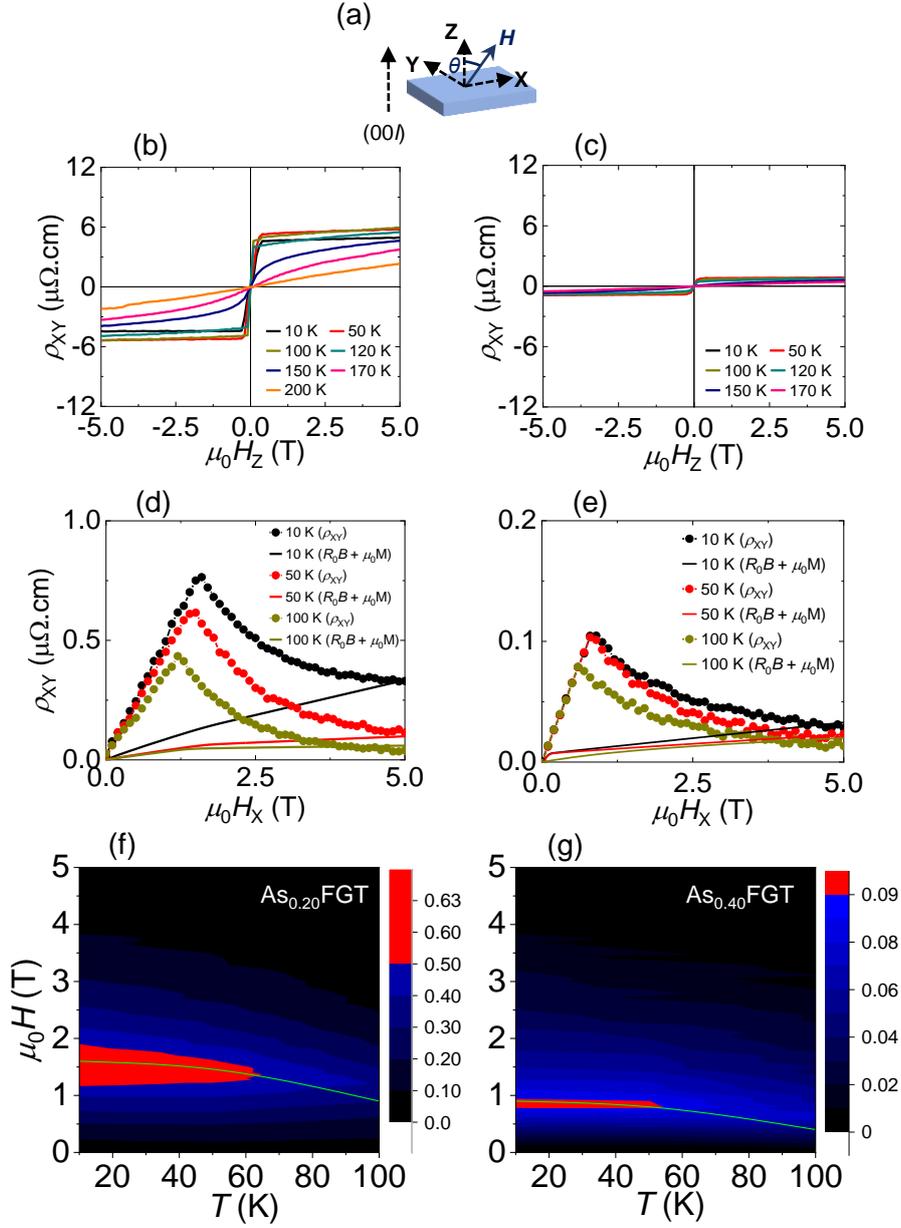

FIG. 4. (a) Hall resistivity ($\rho_{XY}$) versus applied $H_X$ ($\perp$ c-axis $\parallel I$) for As$_x$FGT single-crystal at various temperatures. (b), (c) Temperature dependence of transverse resistivity ($\rho_{XY}$) versus $H_Z$ ($\parallel$ c-axis) for As$_{0.20}$FGT and As$_{0.40}$FGT, respectively. (d), (e) Temperature dependence of $\rho_{XY}$ versus $H_X$ ($\perp$ *c*-axis) for As$_{0.20}$FGT and As$_{0.40}$FGT, respectively. Solid lines in (d), (e) indicate the total contributions from ordinary and anomalous Hall effect contributions, calculated using Eq. (1). (f), (g) Contour mapping of extracted unconventional Hall effect contribution ($\rho_{XY}^U$) as a function of applied magnetic field ($H$) and temperature ($T$) As$_{0.20}$FGT and As$_{0.40}$FGT, respectively, respectively. Yellow solid lines in (f), (g) corresponds to the magnetic field strength where a maximum in $\rho_{XY}^U$ was obtained at a fixed $T$.

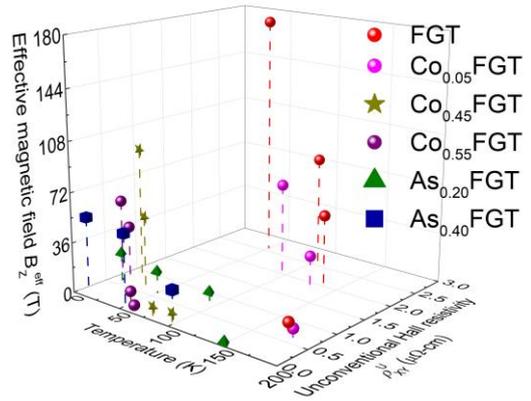

FIG. 5. Three-dimensional temperature ($T$), unconventional Hall resistivity ($\rho_{XY}^U$), and effective emergent magnetic field ($B_Z^{\text{eff}}$) phase diagram for reference FGT doping at the magnetic site (Fe) of FGT [26], and doping at the non-magnetic (Ge) site (this work).

(1), respectively. Figures 4(f),(g) show the $T$-$H$ phase diagram of the extracted $\rho_{XY}^U$, for both the single crystals. Interestingly, an unconventional Hall resistivity ($\rho_{XY}^U$) contribution is also present with doping at the non-magnetic (Ge) site, whose strength is significantly modified with As doping concentration ($x$). In accordance to our previous studies on doping at the magnetic (Fe) site [26], the strength of $\rho_{XY}^U$ and the magnetic field at $\rho_{XY}^U$ maximum monotonically decreases with increasing $T$. Contrary to most of the previous results which have focused on doping effects at the magnetic (Fe) site, our results of an unconventional Hall effect behavior and its modification by doping at the non-magnetic (Ge) site deepens our understanding of emergent topological and/or magnetic phenomenon in 2D vDW magnetic materials, and offers a novel route towards the realization of complicated spin textures in uniaxial vdW FMs, promising from non-collinear magnetism and future spintronic devices.

Finally, we have estimated the strength of the internal emergent magnetic field and discussed the prospect of doping at the non-magnetic (Ge) site versus doping at the magnetic site (Fe) for the vdW FM FGT. Considering a smoothly varying magnetization texture in $As_xFGT$, possibly attributed to the realization of sub-micrometer scale spin structures, the unconventional contribution to Hall effect under the adiabatic limit can be expressed as [42,43]

$$\rho_{XY}^U \approx P\,R_0\,B_Z^{eff} \qquad (2)$$

where, $P$ is the spin polarization of the itinerant electrons and $B_Z^{eff}$ corresponds to the effective magnetic field from the spin texture. An estimation of $P$ has been deduced from the ratio of the ordered moment per magnetic atom (from *M-H* measurements) to the saturated magnetic moment (from Curie-Weiss fitting of *M-T* curves) [42]. Figure 5 shows the obtained values of $B_Z^{eff}$ for doping at the non-magnetic (Ge) site along with a comparison with doping at the magnetic (Fe) site and parent FGT, where $B_Z^{eff}$ for FGT and Co-doped FGT have been calculated from our previous report [26]. The effective magnetic field $B_Z^{eff}$ is ≈ 174 T (at 50 K) for FGT, progressively decreases with doping either at the magnetic (Fe) or non-magnetic (Ge) site, and decreases with increasing $T$. Since the magnitude of $B_Z^{eff}$ is related to the topology of the underlying spin configurations through the density of the texture, a variation of $B_Z^{eff}$ is indicative of a modification in the underlying interactions through doping in the vdW FM FGT. In our previous work, the variation of $B_Z^{eff}$ with magnetic (Fe) site doping was attributed to a modification of underlying complicated magnetic configuration resulting in the formation of aggregate of topological skyrmion bubble-like structures and magnetic domain walls in a stripe domain background [26]. However, for doping at the magnetic (Fe) site and/or decreased Fe concentration, a qualitative and/or quantitative understanding of the underlying interactions with increasing $x$ is much complicated due to interplay of several effects such as lattice expansion (contraction) along the in-plane (out-of-plane) directions [10,26], random substitution of dopant (Co) at $Fe^{2+}$ and/or $Fe^{3+}$ site [25]. On the other hand, the pathway of controlling the effective magnetic field ($B_Z^{eff}$) by doping at the non-magnetic (Ge) site presents a much simpler situation. Here, instead of a random distribution of the dopant at the different Fe sites, it provides a preferably cleaner way of tuning the magnetic and magnetotransport properties while leaving the crystal structure intact. We observe that As-doping at the Ge site is associated with lattice contraction (expansion) along the in-plane (out-of-plane) directions [27], opposite behavior compared to the magnetic (Fe) site doping case [26]. Considering the frustrated magnetic configuration of FGT under the application of magnetic fields ($H_X \parallel I \perp c$-axis), the reduction of a $\rho_{XY}^U$ and $B_Z^{eff}$ with increasing As doping might originate from modification of magnetic interaction strength due to a change in Fe-Fe bond lengths. We believe that our results might serve as a starting point for future theoretical and experimental studies to investigate the complicated underlying magnetic configurations, down to the monolayer limit. More importantly, the observed results of a variation of $B_Z^{eff}$ with doping at the non-magnetic (Ge) or magnetic (Fe) site entails flexibility towards material design with tailored micromagnetic (such as $M_S$, $K_{eff}$) properties. The present study offers an unexplored pathway towards controlling of the underlying interactions necessary for the realization of unconventional spin textures, crucial for the realization of vdW-based spintronic devices for conventional and unconventional computing paradigms.

## IV. CONCLUSIONS

In conclusion, we have investigated the effect of As doping at the non-magnetic (Ge) on the magnetic and magnetotransport properties of uniaxial vdW FM $Fe_3GeTe_2$. Temperature-dependent magnetization measurements reveal a modification of the transition temperature with increasing As concentration, and a significant reduction of the uniaxial magnetic anisotropy, compared to parent FGT. Interestingly, magnetotransport measurements show the existence of an unconventional Hall effect whose strength is significantly modified with As-doping concentration ($x$). The present experimental results provide considerable insights into the origin of unconventional Hall effect behavior in vdW FM and suggest a possible route towards the realization of stable topological spin textures, prospective for vdW-based spintronic physics and applications.

## V. ACKNOWLEDGEMENTS

The authors thank H. Ohno for fruitful discussions and valuable comments. R. Roy Chowdhury acknowledges Department of Science and Technology (DST), Government of India, for financial support (Grant no. DST/INSPIRE/04/2018/001755). R. P. S. acknowledges Science and Engineering Research Board (SERB), Government of India, for Core Research Grant CRG/2019/001028. A portion of this work was supported by JSPS Kakenhi 19H05622, 20K15155 and RIEC International Cooperative Research Projects, Tohoku University.

## APPENDIX A: ANOMALOUS HALL CONDUCTIVITY IN $Fe_3GeTe_2$ AND AS-DOPED $Fe_3GeTe_2$

The presence of spin-orbit interaction in FGT has been shown to stabilize topological nodal lines, manifesting in a large Berry curvature leading to significantly large anomalous Hall conductivity (AHC) ($\sigma_{XY}$) and anomalous Hall angle ($\theta_{AHA}$) [14]. To clarify the origin of the anomalous Hall effect in undoped and As-doped FGT, we plot the $\sigma_{XY}$ versus longitudinal conductivity ($\sigma_{XX}$) for FGT and its As-doped counterparts ($As_{0.20}FGT$, and $As_{0.40}FGT$) (Fig. 6). Interestingly, even with significant As-doping levels, we observe a large $\sigma_{XY}$, much larger than 3D ferromagnets (FMs) (such as Fe [44]) or archetype chiral magnets (such as MnSi [45], $Co_3Sn_2S_2$ [46]), and to the quantum conductance per atomic layer. More interestingly, for both FGT and its As-doped counterparts, $\sigma_{XY}$ is virtually independent of $\sigma_{XX}$, indicating an intrinsic Berry curvature origin of the anomalous Hall effect.

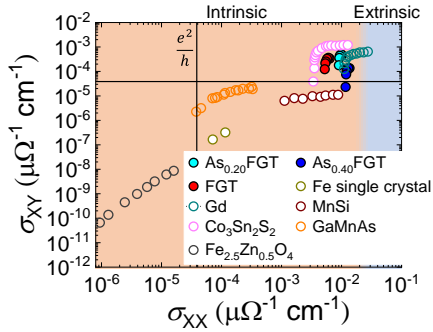

FIG. 6. Anomalous Hall conductivity ($\sigma_{XY}$) versus longitudinal conductivity ($\sigma_{XX}$) for Fe single crystals [44], Gd thin film [44], chiral magnet MnSi [45], $Co_3Sn_2S_2$ [46], ferromagnetic semiconductor GaMnAs [47], magnetite $Fe_{2.5}Zn_{0.5}O_4$ [48], van der Waals $Fe_3GeTe_2$ [26], and As-doped FGT (this work). The black horizontal and vertical line denotes the expected value for quantum Hall effect per atomic layer ($e^2/h$) ($e$ = electronic charge, $h$ is Planck's Constant).

## APPENDIX B: EVALUATION OF DOMAIN WIDTH

For the case of complicated domain patterns such as those observed for vdW FMs, the conventional techniques for the determination of magnetic domain and DW width is not applicable, leading us to utilize a stereological method. Figure 7(a) shows the MOKE image of ZFC remnant magnetic domain configuration for FGT (at 100 K). Random lines of various lengths (denoted as test line) were drawn across the domain pattern at different positions (indicated by lines A and B). The number of intersections of each line with a DW equals the number of sudden rise or drop of pixel intensity from a low value (~ 0) to a high value (~ 255) and vice versa (Fig. 7(b)). The average magnetic domain size ($D$) was determined as

$$D = \frac{2 \times \text{Total test line length}}{\pi \times \text{Number of intersections}}$$

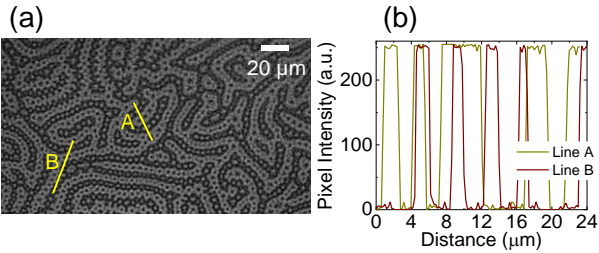

FIG. 7. (a) Differential p-MOKE image of remnant magnetic state for FGT, respectively at 100 K. Black (grey) areas correspond to magnetization pointing in parallel (anti-parallel) direction with respect to *c*-axis (out of the plane of paper). The test lines A and B (yellow lines) were drawn to evaluate the domain period. (b) Pixel intensity versus test line length for the yellow lines in the previous figure. The minimum pixel intensity (~ 0) correspond to the regions with the magnetization orienting antiparallel to *c*-axis while the maximum pixel intensity (~ 255) correspond to the magnetization along the *c*-axis.

## APPENDIX C: LONGITUDINAL RESISTIVITY

Figure 8 shows the temperature dependent zero-field longitudinal resistivity of $As_xFGT$ ($x$ = 0.20, 0.40) and parent FGT, measured using a four-probe geometry, and exhibits metallic behavior for all the single crystals investigated in this study. For both $As_xFGT$ and parent FGT, a reduction of $T$ from 300 K results in decrease of resistivity, showing a hump-like feature close to $T_C$ (determined from magnetization measurements). Note that the As-doping results in a considerable reduction of the resistivity, consistent with previous results on polycrystalline samples. These values of resistivities were used to determine the anomalous Hall contribution of Hall resistivity (Eq. (2)).

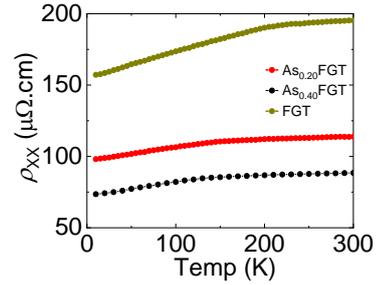

FIG. 8. Longitudinal resistivity ($\rho_{XX}$) versus temperature ($T$) for $As_{0.20}FGT$, $As_{0.40}FGT$, and parent FGT single-crystalline samples.